\documentclass[a4paper,11pt,unpublished,accepted=2022-04-26]{quantumarticle}
\pdfoutput=1

\usepackage{graphicx}
\usepackage[numbers,sort&compress]{natbib}
\usepackage[utf8]{inputenc}
\usepackage[english]{babel}
\usepackage[T1]{fontenc}
\usepackage{amsmath}
\usepackage[colorlinks=false]{hyperref}
\usepackage{dsfont}
\usepackage{balance}
\usepackage{flushend}

\begin{document}

\title{Understanding and Improving Critical Metrology. Quenching Superradiant Light-Matter Systems Beyond the Critical Point}

\author{Karol Gietka}
\orcid{0000-0001-7700-3208}
\email{karol.gietka@oist.jp}
\author{Lewis Ruks}
\orcid{0000-0002-5593-3233}
\author{Thomas Busch}
\orcid{0000-0003-0535-2833}
\address{Quantum Systems Unit, Okinawa Institute of Science and Technology Graduate University, Onna, Okinawa 904-0495, Japan}


\begin{abstract}
We carefully examine critical metrology and present an improved critical quantum metrology protocol which relies on quenching a system exhibiting a superradiant quantum phase transition beyond its critical point. We show that this approach can lead to an exponential increase of the quantum Fisher information in time with respect to existing critical quantum metrology protocols relying on quenching close to the critical point and observing power law behaviour. We demonstrate that the Cram\'er-Rao bound can be saturated in our protocol through the standard homodyne detection scheme. We explicitly show its advantage using the archetypal setting of the Dicke model and explore a quantum gas coupled to a single-mode cavity field as a potential platform. In this case an additional exponential enhancement of the quantum Fisher information can in practice be observed with the number of atoms $N$ in the cavity, even in the absence of $N$-body coupling terms.
\end{abstract}
\maketitle


\section{Introduction}
Controlling and manipulating light-matter interactions is currently one of the most rapidly growing and evolving field of physics~\cite{polzik2010lightmatterreview}. Of particular interest is the coupling of a cavity field to a gas of atoms---quantum-gas cavity quantum electrodynamics---where the interaction between light and matter is greatly enhanced due to the light passing through the same atomic system many times. Whilst the back-action of the atoms on the light field is typically neglected in free-space configurations, its necessary inclusion in cavity quantum electrodynamics leads to complex nonlinear coupled dynamics~\cite{esslinger2031cavityqed}. The rich ensuing dynamics in turn open up possibilities to use these systems to simulate fundamental solid-state physical systems and explore non-equilibrium many-body phenomena beyond the scope of conventional condensed matter systems~\cite{mivehvar2021cavity}. Moreover, due to the presence of non-classical correlations and the feasibility of non-destructive monitoring of these systems through photons leaking from the cavity mirrors, they serve as an ideal platform for precise measurements of unknown physical parameters beyond the standard quantum limit~\cite{2007cavityqedheisenberg,2011GammelmarkDickeISING,2016PTcavity,gietka2015qednonclass,2019quantummetrologycavityqed,gietka2019ssbgrc,2020precisecavity,Gietka_2021njpcavitymagneto,chu2021quantum}. Therefore, quantum-gas cavity quantum electrodynamics systems may lead to the development of new paradigms in quantum metrology.

Quantum metrology \cite{giovannetti2004quantumenhanced} is a framework in which quantum-mechanical effects \cite{giovannetti2006quantummetrology}, for example,  non-classical correlations  \cite{braun2013parestisinglmode} and quantum entanglement \cite{pezze2009entanglementhL}, are being used to enhance the precision of measurement beyond the standard quantum limit \cite{giovannetti2006quantummetrology}. The figure of merit in quantum metrology is the quantum Cram\'er-Rao bound \cite{2007PRL_GeneralizedLimits}, which sets the ultimate limit of measurement precision of an unknown parameter $\lambda$ using limited resources---such as time $t$ and number of particles---that can be potentially attained: $\Delta^2 \lambda \geq {1}/{{m \mathcal{I}_\lambda}}$. Here $m$ is the number of estimation protocol repetitions and $\mathcal{I}_\lambda$ is the quantum Fisher information, which for pure states takes the form \cite{caves1994geomqs} $\mathcal{I}_\lambda = 4(\langle \partial_\lambda \psi|\partial_\lambda \psi \rangle - \langle \partial_\lambda \psi|\psi \rangle^2)$. From the practical point of view, an important quantity is the classical Fisher information $\mathcal{F}_\lambda = \sum_\xi  \frac{1}{p(\xi|\lambda)}[\partial_\lambda p(\xi|\lambda)]^2$. This quantity takes into account the performed measurement through the corresponding conditional probability distribution  $p(\xi|\lambda)$ and can be equal to its quantum version only for the optimal choice of measurement. However, in certain cases the optimal measurement might be beyond the scope of the experimental possibilities.

\begin{figure}[tb!]
    \centering
    \includegraphics[width=\columnwidth]{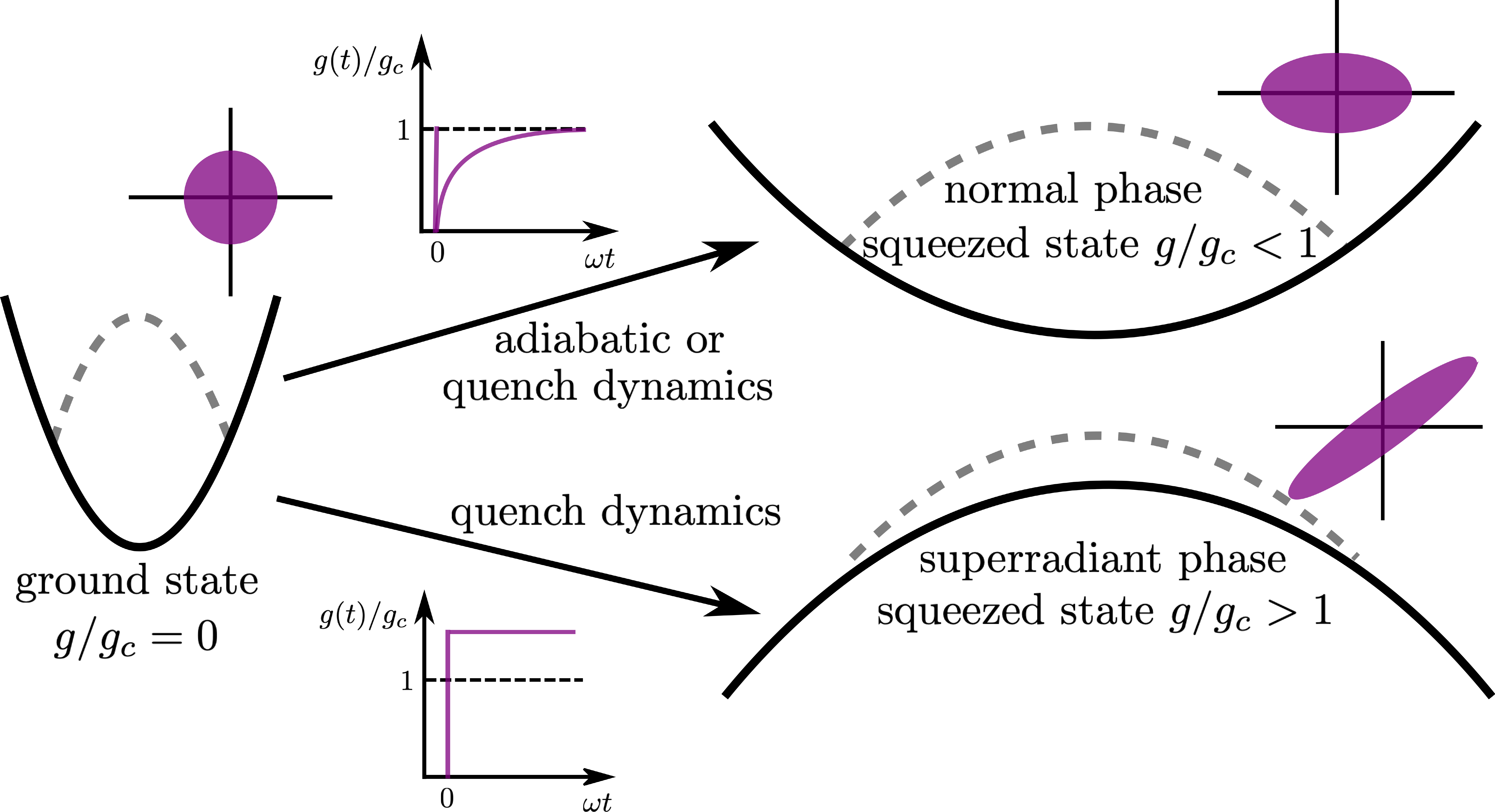}
    \caption{The schematic represents a toy model for a quantum phase transition ($g/g_c>1$ is the superradiant phase), with the black line being the effective potential that is felt by a quantum state (dashed-gray line). Driving the system close to the critical point ($g/g_c \sim 1)$ creates the correlated (squeezed) excitations at a very slow rate (critical slowing down), since the effective potential is still of trapping form. However, if the system is quenched beyond the critical point ($g/g_c > 1$), the same number of correlated excitations can be generated much faster since the initial state will behave as if it was placed in an inverted harmonic oscillator potential. The purple ellipses represent the phase space picture of the state.}
    \label{fig:schmeatic}
\end{figure}

In a general metrology scheme, we can consider a Hamiltonian of the form ($\hbar =1$ in the following)
\begin{equation}
    \hat H_\mathrm{QM} = \lambda \hat H_\lambda + \hat H_{\mathbf{x}}(\mathbf{x}),
\end{equation}
where $\hat H_\lambda$ is responsible for imprinting the information about the unknown parameter $\lambda$ into the initial state $|\psi_0 \rangle$, and $\hat H_{\mathbf{x}}(\mathbf{x})$ describes other dynamics which in general can depend on a set of $n$ parameters $\mathbf{x} = x_1,x_2,\ldots,x_n$. Provided that one has full control over $\mathbf{x}$, quantum metrology protocols typically first employ $\hat H_{\mathbf{x}}(\mathbf{x})$ to prepare a suitable (for example, entangled) initial state, after which $\hat H_{\mathbf{x}}(\mathbf{x})$ is set to 0. Subsequently the information about the unknown parameter is imprinted on the prepared initial state through the unitary evolution operator $\hat U(\lambda,t) = \exp(- i t \lambda \hat H_\lambda)$.
By initially preparing a superposition of the minimal and the maximal eigenvalue eigenstates of the so-called local generator 
$\hat h_\lambda = i\hat U^\dagger\partial_\lambda\hat U$ (which in this case is $\hat h_\lambda = \hat H_\lambda$), it is possible to maximize the quantum Fisher information and reach the so-called Heisenberg limit of the sensitivity. 
Note that it can be also shown that the quantum Fisher information can be calculated as the variance of the local generator on the initial state $\mathcal{I}_\lambda = 4 \Delta^2 \hat h_\lambda$. For example, in $N$ two-mode atomic systems the Heisenberg limit is $\Delta^2 \lambda = 1/(Nt)^2$~\cite{smerzi2018reviewatomquantum}, attained by preparing the maximally entangled state. Here $t$ is the phase acquisition time (note that the initial state preparation time is typically not taken into account). For frequency estimation with a single-mode Gaussian field, the optimal sensitivity becomes $\Delta^2 \lambda = 1/8t^2(\langle n\rangle^2+\langle n\rangle)$~\cite{Polino2020photonicqm} ($\langle n \rangle$ being the average number of photons) and is saturated using a squeezed vacuum state. If the imprinting mechanism $\hat H_\lambda$ is additionally time-dependent, optimal control techniques can be used to maximize the quantum Fisher information by changing $\mathbf{x}$ in time such that the instantaneous state becomes the superposition of maximal and minimal eigenvalue eigenstate of the local generator~\cite{2017timedependentmetrologypang,2021superheisenerandheisenberg}. However, the Hamiltonian might then be a complicated function of time as it is no longer composed of only $ \hat H_\lambda$ but also involves $\hat H(\mathbf{x})$ which in general does not commute with $\hat H_\lambda$. Note that if the information about the unknown parameter is imprinted non-linearly, i.e., with a $k$-body term $\hat H_\lambda^{(k)}$, the power-law scaling with $N$ in the Heisenberg limit is modified. For example, in a system composed of $N$ two-mode systems with a multi-body coupling of degree $k$, the Heisenberg limit becomes $\Delta^2 \lambda = 1/(N^{k/2}t)^2$ \cite{2007PRL_GeneralizedLimits}. In principle, by exploiting $N$-body interactions, it could be possible to obtain an exponential enhancement of achievable precision~\cite{2008Exponetially_enhanced_quantum_metrology_PRL}, however, it has been claimed that such interactions are \emph{unphysical}~\cite{2007PRL_GeneralizedLimits}.

Although metrological schemes such as those described can in principle be used to reach the Heisenberg limit, they are often extremely demanding from an experimental point of view. The preparation of maximally entangled states by itself is already a challenging task---especially for systems composed of many constituents---which is complicated further by manipulation and measurement. Additionally, highly non-classical states are also extremely fragile to the effect of decoherence. Heisenberg-limited metrology is therefore currently restricted to proof-of-principle experiments, often with systems composed of only a few elements such as atoms or photons~\cite{marciniak2021optimal}. However, in many cases the standard quantum limit can be overcome with the use of less exotic states such as spin-squeezed states of atoms \cite{smerzi2018reviewatomquantum} or squeezed states of photons \cite{Polino2020photonicqm}. Finally, the specific measurement required to saturate the Cram\'er-Rao bound might not be possible in certain cases; for example, the optimal measurement might correspond to a projection onto a highly entangled state.  

A simpler metrological protocol can be constructed for situations where one does not have full control over $\mathbf{x}$. In this case, one can imprint the information about the unknown parameter $\lambda$ and build up the quantum correlations simultaneously by starting from an initially uncorrelated state \cite{rossi2020noisyenhancednondemo}. This approach includes explicitly the state preparation time in the metrological protocol in contrast to protocols in which an optimal initial state is already prepared.  However, the cost of reducing the complexity of the metrological protocol is the inability to maximize the quantum Fisher information and reach the Heisenberg limit~\cite{2021gietka_adiabatic}. Nevertheless, it is still possible to beat the standard quantum limit with moderately non-classical states, and in certain cases reach the Heisenberg scaling. That is to say, the sensitivity follows the same power-law scaling with time and number of particles as the Heisenberg limit, but contains a constant prefactor $f >1$, i.e., $\Delta^2 \lambda  \geq f/N^2 T^2$. In certain cases, whilst the sensitivity exhibits Heisenberg scaling, it can still fail to beat the standard quantum limit of precision if $f=N_f>N$. In principle one could then only beat it for extremely large system sizes which might go beyond the scope of experimental realizations~\cite{braun2018withouthent}.

A key example of simultaneously imprinting the information about the unknown parameter whilst creating quantum correlations is the so-called critical quantum metrology~\cite{Zanardi2008QuantumCriticality,paris2008spincriticality,porras2013adiabaticqm,tsang2013qted,paris2014qmLMG,macieszczak2016dpt,paris2016dickqpt,porras2017quantumsensingcriticality,chitra2019qtransducerddpt,felicetti2020criticalqm,Salado_Mej_a_2021,hatomura2021symmetryprotected,ilias2021criticality,garbe2021critical,gietka2021squeezing}, which has attracted much attention in recent years. Critical quantum metrology takes advantage of extreme sensitivity of the quantum state to perturbations when near a quantum phase transition. When changes in the unknown parameter effect such perturbations, one can use this extreme sensitivity for high-precision measurements. If the Hamiltonian exhibits a quantum phase transition at  $\mathbf{x}_c$, one can use the critical quantum metrology approach to adiabatically drive the system near this critical point and perform a suitable measurement yielding a large Fisher information. This protocol is effective because in the thermodynamic limit of continuous quantum phase transitions, the energy gap above the ground state closes \cite{sachdev2007quantum}. The effect of a vanishing energy gap can  be explicitly seen if we calculate the quantum Fisher information for the ground state $|\psi_0(\lambda)\rangle$ of the Hamiltonian $\hat H(\lambda) = \sum_{n=0}E_n(\lambda)|\psi_n(\lambda)\rangle\langle\psi_n(\lambda)|$ \cite{you2007fidelitycriticality}
\begin{equation}
    \mathcal{I}_\lambda = 4 \sum_{n\neq 0} \frac{|\langle \psi_n(\lambda) |\partial_\lambda \hat H(\lambda)| \psi_0(\lambda)\rangle|^2}{[E_n(\lambda)-E_0(\lambda)]^2}.
\end{equation}
From this expression, it is clear that the quantum Fisher information diverges through the denominator if the energy gap above the ground state closes. This property may lead, in principle, to an arbitrarily high estimation precision in the thermodynamic limit. Unfortunately, the adiabatic theorem states that a quantum system remains in its ground state only if it is varied slowly enough such that no excitations occur~\cite{1928Adiabatic}; a sufficient condition is that the time time-scale of variation far exceeds the inverse energy gap $t \gg 1/\Delta E.$ Therefore, close to the critical point the metrological protocol has to be infinitely slow (\emph{critical slowing down}) as the energy gap closes. This means that from the viewpoint of metrology the divergence of the quantum Fisher information is actually due to the diverging time required for the adiabatic protocol~\cite{Rams2018atthelimitsofcriticality,felicetti2020criticalqm}.  Moreover, since the protocol relies on creating quantum correlations generated during the adiabatic time evolution in the vicinity of the critical point, it also means that this protocol cannot reach the Heisenberg limit \cite{2021gietka_adiabatic}. Therefore, adiabatic quantum metrology protocols increase dramatically the time of a single measurement at the cost of not reaching the Heisenberg limit.  Additionally, such a protocol might very likely operate on time scales larger than decoherence times and will also be unable to reach the Heisenberg limit in this respect. Recently, it has been shown that in certain cases adiabatic time evolution is not required and metrology that can lead to the Heisenberg scaling can be done using quenches to or close to the critical point \cite{chu2021dynamiccriticalmetrology}. However, such dynamical critical quantum metrology protocols have operated within in a single phase of the system and also require extremely large time-scales.

In systems exhibiting a superradiant phase transition, the exponential growth of correlated excitations observed in a quench across the critical point~\cite{invertedoscillator2021} constitutes a macroscopic degree of freedom with a large potential for information storage that has not yet been explored. Quenching across a transition (see Fig.~\ref{fig:schmeatic}) further allows one to sidestep the issue of critical slowing down encountered when preparing states near a critical point (i.e., adiabatic critical quantum metrology). Therefore, an alternative aim of quantum metrology in systems exhibiting a superradiant phase transition could be to quench far beyond the critical point and imprint information into the macroscopically populated field (however, the coupling to highly sensitive systems such as biological samples may not be suitable for such a protocol~\cite{PerarnauLlobet2021weaklyinvasive}). Given the prevalence of superradiant phase transitions in light-matter systems~\cite{PhysRevLett.124.040404}, we expect that the protocol we present in the following will be applicable to a wide range of scenarios.

In the following we explore the dynamical approach to quantum metrology with light-matter systems exhibiting superradiant quantum phase transitions, which we show leads to an exponentially growing quantum Fisher information caused by an exponential growth correlated photons on an arbitrarily fast time scale (scaling inversely with the coupling strength after the quench). That said, the exponential growth is limited until the critical amount of excitations (photons) is reached as will explain in the following Sections. In contrast to the existing critical metrology proposals, our approach relies on creating both correlations and excitations by quenching beyond the critical point. We will demonstrate this in practice using the example of the paradigmatic Dicke model and show that quadrature measurements can saturate the Cram\'er-Rao bound (extending existing results in the normal or superradiant phase), producing an exponential growth of Fisher information in time. This is in contrast to existing proposals in critical quantum metrology that result in a power-law scaling with time, with a recent result obtaining a $t^4$ scaling~\cite{felicetti2020criticalqm}. 

In addition, we propose a gas of atoms coupled to a single-mode cavity field as a potential test platform for our protocol. We demonstrate that when the requirement of extensivity in system quantities is dropped, the natural light-matter coupling Hamiltonian results in a quantum Fisher information scaling as $I_{\lambda} \sim \exp(\alpha \sqrt{N}t)$ after a quench into the superradiant regime where self-organization of the atomic gas is observed. This illustrates that an exponential enhancement of Fisher information is observable not only in time, but also in practice with $N$, despite the system only having two-body interactions~\cite{2007PRL_GeneralizedLimits}. The exponential growth of the quantum Fisher information is also present in finite-component systems (thus finite-component quantum phase transitions).


\section{Dicke model}
Superradiant quantum phase transitions are a ubiquitous phenomena in quantum optics~\cite{PhysRevLett.124.040404} which can occur in a collection of two-level systems (typically atoms) interacting with a single harmonic oscillator (electromagnetic field). The superradiant phase is characterised by a macroscopic number of excitations (photons) in the twofold degenerated ground state~\cite{Larson_2017,PhysRevA.100.063820}. In order to illustrate how our metrology scheme exploits the superradiant phase transition, let us consider the Dicke model. This well-known model describes the interaction of $N$ two-level particles with a single-mode field and is known to exhibit a superradiant phase transition. Using the collective spin operators $\hat S_i = \sum_{i=1}^{N}\hat \sigma_i/2$ with $\hat \sigma_i$ being the $i$th Pauli matrix, the Dicke model can be expressed as
\begin{equation}\label{eq:qr}
    \hat H_{\mathrm{DM}} = \omega \hat a^\dagger \hat a + {\Omega} \hat S_z +\frac{g}{\sqrt{N}}\left( \hat a^\dagger + \hat a \right) \hat S_x,
\end{equation}
where $\omega$ is the frequency of the bosonic field represented by its creation and annihilation operators $\hat a^\dagger$ and $\hat a$, $\Omega$ is the energy splitting of a two-level particle represented by Pauli matrices $\hat \sigma_i$, and $g$ is the coupling parameter between the bosonic field and the two-level particle. We note that the ${1}/{\sqrt{N}}$ term is in practice introduced to guarantee extensivity of system quantities. The Dicke model then exhibits a quantum phase transition in the limit of $\sqrt{\omega/\Omega N} \rightarrow 0$ for the critical coupling strength defined as $g_c\equiv \sqrt{\omega \Omega}$. In the thermodynamic limit $N \to \infty$, the quantum phase transition occurs for an arbitrary but finite $\sqrt{\omega/\Omega}$. For the other limiting case, i.e., $N=1$ (quantum Rabi model), the (finite component) quantum phase transition occurs for $\sqrt{\omega/\Omega} \rightarrow 0$, which becomes an equivalent of the thermodynamic limit \cite{plenio2015rabimodelqpt}. To simplify the Dicke model we can apply the Schrieffer-Wolff transformation~\cite{2011schriefferwolff} $ \hat U_{\mathrm{SW}}\hat H_{\mathrm{QRM}} \hat U_{\mathrm{SW}}^\dagger$ with $\hat U_{\mathrm{SW}} =\exp\{i (g/\Omega\sqrt{N}) (\hat a^\dagger + \hat a)\hat S_y\}$. In the limit of $\sqrt{\omega/\Omega} \rightarrow 0$ the transformation is exact and yields 
\begin{equation}\label{eq:effqr}
    \hat H_{\mathrm{DM}} \simeq \omega \hat a^\dagger \hat a + {\Omega} \hat S_z +\frac{g^2}{2\Omega N}\left( \hat a^\dagger + \hat a \right)^2 \hat S_z.
\end{equation}
This effective model allows us to analytically calculate the spectrum including the critical ground state and thus the quantum Fisher information. The eigen-states of the Dicke model under the Schrieffer-Wolff transformation become (for clarity we restrict to the low energy sector)
\begin{equation}
    |\psi_n \rangle = \hat S(\xi)| n \rangle \otimes |\!\downarrow\, \rangle,
\end{equation}
where $\hat S(\xi) \equiv \exp\{(\xi/2)(\hat a^\dagger)^2-(\xi^*/2)\hat a^2\}$ is the squeezing operator with $\xi = -\frac{1}{4} \ln\{1-(g/g_c)^2\}$ the squeezing parameter which is real only for $g < g_c$ and $|\! \downarrow \,\rangle$ is the collective spin down state. The condition $g<g_{c}$ restricts the validity of the ground state to the normal phase. Nevertheless, as has been recently shown the effective Hamiltonian can be valid dynamically given arbitrary $g$ for a limited time~\cite{invertedoscillator2021} which in the thermodynamic limit becomes infinity. The quantum Fisher information with respect to $\lambda$--- (which can be set to either $\omega$ or $\Omega$)---for the $n$th excited state of the Dicke model $|\psi_0 \rangle  = \hat S(\xi)|n\rangle \otimes |\!\downarrow\, \rangle$ becomes (see the Appendix \ref{app:qfinormal})
\begin{equation}
    \mathcal{I}_\lambda^a = \frac{1+n+n^2}{ 8\lambda^2  \left(1 - \frac{g ^2}{g_c^2} \right)^2}\frac{g ^4}{g_c^4}
\end{equation}
which is clearly divergent for $g \simeq g_c$ (superscript $a$ implies adiabatic dynamics). However, in practice there is an implicit dependence on time and when this dependence is made explicit the divergence disappears~\cite{felicetti2020criticalqm}. This happens because the time required to prepare a critical ground state diverges as the reciprocal of the energy gap, [given as the square root of $\delta_\epsilon = 4 \omega^2(1-g^2/g_c^2)$], near the critical point at the quantum phase transition.

In what follows, we will show how to eliminate the problem of critical slowing down while still exploiting criticality.


\begin{figure*}[htb!]
    \centering
    \includegraphics[width=\textwidth]{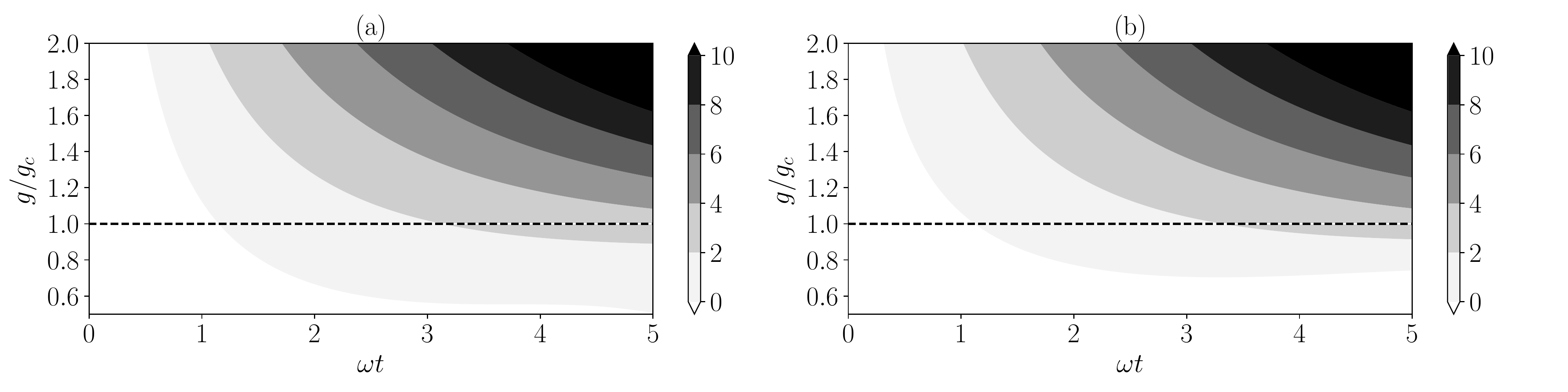}
    \caption{The logarithm of the quantum Fisher information (for initial vacuum state) normalized to $\lambda^2$ as a function of $g/g_c$ and time expressed in the units of $\omega^{-1}$. Panel (a) and (b) depict the  quantum Fisher information for $\lambda = \omega$ and $\lambda = \Omega$, respectively. The dashed line illustrates the quantum Fisher information attainable by quenching the system to the critical point.}
    \label{fig:fig1}
\end{figure*}

\begin{figure*}[htb!]
    \centering
    \includegraphics[width=0.99\textwidth]{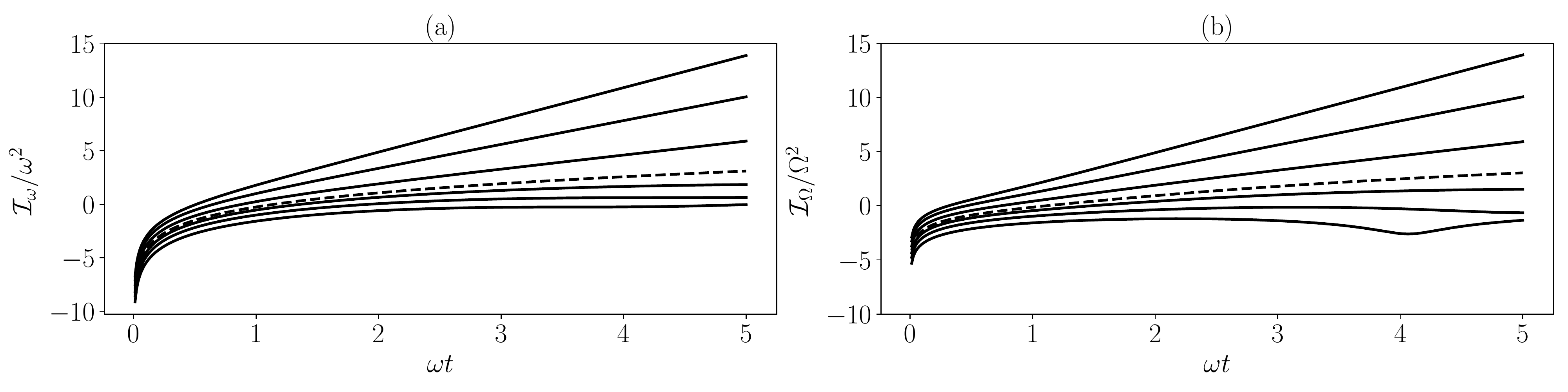}
    \caption{Extension of Fig.~\ref{fig:fig1} showing slices of the logarithm of the quantum Fisher information for different values of $g/g_c$ (from the bottom line $0.5,0.68,0.87,1,1.25,1.62,2$). The dashed line represents $g/g_c=1$.}
    \label{fig:fig1b}
\end{figure*}
 
\section{Quenching beyond the critical point}
Although the effective Hamiltonian from Eq.~\eqref{eq:effqr} cannot be used to describe the ground state of the system in the superradiant phase, it can be still used to describe dynamics of a quench from any $g < g_{c}$ to $g > g_c$ provided that~\cite{invertedoscillator2021}
\begin{equation}\label{eq:initialstatecondition}
     \frac{1}{4} \frac{g^2 }{g_c^2 }\frac{ \omega}{\Omega}\ll 1,
\end{equation}
and for a limited time related to the number of photons in the ground state of the quenched Hamiltonian
\begin{equation}\label{eq:photonnumbercondition}
    \left\langle \hat a^\dagger \hat a \right\rangle_t < \frac{1}{32} {\frac{N \Omega }{ \omega }}\left({\frac{g^2}{g_c^2}-\frac{g_c^2}{g^2}}\right),
\end{equation}
which in the thermodynamic limit $\sqrt{N\Omega/\omega}\rightarrow \infty$ becomes infinity  (see Ref.~\cite{invertedoscillator2021} for a detailed derivation). Therefore, for the purpose of analytical analysis, we assume now that Eq.~\eqref{eq:initialstatecondition} and Eq.~\eqref{eq:photonnumbercondition} are satisfied so the use of the effective Hamiltonian is justified. 

In order to shed some light on the system dynamics we rewrite the transformed Dicke Hamiltonian as (see Appendix~\ref{app:guidline})

\begin{equation}
\begin{split}
  \hat H_{\mathrm{DM}} =&\left(\omega + \frac{g^2}{\Omega N}\hat S_z \right) \hat a^\dagger \hat a + {\Omega} \hat S_z \\&+\frac{g^2}{2\Omega N}\left( \hat a^{\dagger2} + \hat a^2 \right) \hat S_z
  \end{split}
\end{equation}
The initial spin state $|\!\downarrow\, \rangle$ is an eigenstate of $\hat S_z$ (ground state for $g/g_c=0$). Since $[\hat{H}_{\text{DM}},\hat{S}_{z}]=0,$ we replace $\hat S_z \to -N/2$ and drop terms proportional to identity to obtain a purely photonic and quadratic Hamiltonian commonly appearing in Gaussian metrology~\cite{2012Gaussianreview} and is essential to understand critical metrology:
\begin{equation}\label{eq:invertedoscillatorH}
   \hat H = \left(\omega- \frac{g^2}{2\Omega} \right) \hat a^\dagger \hat a  -\frac{g^2}{4\Omega} \left( \hat a^{\dagger 2} + \hat a^2 \right).
\end{equation}
{ In fact, this Hamiltonian describes a harmonic oscillator in quadrature space with a tunable frequency
\begin{equation}\label{eq:invosc}
   \hat H = \frac{\omega}{2} \hat P^2 + \frac{\omega}{2}\left(1-\frac{g^2}{g_c^2}\right)\hat X^2,
\end{equation}
whose physics is easy to understand. If we quench to $g \leq g_c$ the vacuum starts to expand as the frequency of the harmonic oscillator is decreased. This will happen until the frequency of the effective harmonic oscillator in quadrature space becomes 0~\cite{invertedoscillator2021} which coincides with the critical coupling $g=g_c$. At that point, the initial state is being constantly squeezed at a finite (quadratic) rate in analogy to free particle expansion. By then quenching the interaction parameter further beyond the critical coupling $g_c$, this rate of squeezing increases (from quadratic to exponential)}, leading to a divergence of squeezing over an arbitrarily short time-scale in the thermodynamic limit. That said, for a finite size system the squeezing will last only until the maximum number of excitations that a system can support is reached [see Eq. \eqref{eq:photonnumbercondition}].

It is well known that squeezing Hamiltonian leads to an exponential growth of the number of photons~\cite{1983squeezedstatesreviewnature}. On the other hand, it is also well known that the quantum Fisher information is related to the number of particles~\cite{smerzi2018reviewatomquantum,Polino2020photonicqm}; therefore we can expect exponential growth of the quantum Fisher information as well. This idea is made explicit from a heuristic extension of the analytical expression of the quantum Fisher information derived in the normal phase. In Ref.~\cite{chu2021dynamiccriticalmetrology} it has been shown that if a Hamiltonian $\hat H = \hat H_0 +\lambda \hat H_\lambda$ satisfies the eigenvalue equation 
\begin{equation}
    [\hat H,i\sqrt{\delta_\epsilon} \hat C - \hat D] = \sqrt{\delta_\epsilon}\left(i\sqrt{\delta_\epsilon} \hat C - \hat D \right),
\end{equation}
with $\hat C = -i[\hat H, \hat H_\lambda]$, $\hat D = -i[\hat H,\hat C]$, and $\delta_\epsilon \hat C = [\hat H,-\hat D]/i$, the local generator $\hat h_\lambda$ becomes~\cite{2014Generalhamiltoniantime}
\begin{equation}\label{eq:QFIdick}
    \hat h_\lambda = \hat H_\lambda t +\frac{\cos(\sqrt{\delta_\epsilon}t)-1}{\delta_\epsilon}\hat C -\frac{\sin(\sqrt{\delta_\epsilon}t)-\sqrt{\delta_\epsilon} t}{\sqrt{\delta_\epsilon}\delta_\epsilon}\hat D,
\end{equation}
and the quantum Fisher information can be simply calculated as the variance of the local generator $\mathcal{I}_\lambda = 4 \Delta^2 \hat h_\lambda$.

Let us now first explicitly demonstrate the shortcoming of critical metrology again for the Dicke model~\cite{chu2021dynamiccriticalmetrology}. It can be shown that in the normal phase the energy gap is given as the square root of $\delta_\epsilon = 4\omega^2(1-g^2/g_c^2)$. Therefore, it might naively seem that going to a critical point $g=g_c$ of a quantum phase transition is indeed optimal as for $\delta_\epsilon \rightarrow 0$ near the critical point the quantum Fisher information might diverge~\cite{chu2021dynamiccriticalmetrology}. However, in order to see that divergence of the quantum Fisher information the condition $\sqrt{\delta_\epsilon } t = 2\sqrt{\omega^2(1-g^2/\omega\Omega)} t = \mathcal{O}(1)$ has to be satisfied. This in turn means that in order to see the divergence the time has to satisfy $t \approx 1/\sqrt{\delta_{e}}$ which happens for $t \to \infty$ as $1-g^2/\omega\Omega \simeq 0$. In other words, the divergence caused by the vanishing energy gap requires time approaching infinity~\cite{Rams2018atthelimitsofcriticality,2021gietka_adiabatic}. However, from the discussion above it seems that going beyond the critical point might actually lead to a divergence of the quantum Fisher information on exponentially shorter time scales. Let us therefore analyse further the quantum Fisher information, or specifically the local generator $\hat h_\lambda$. For the Dicke model $\delta_\epsilon = 4\omega^2(1-g^2/\omega\Omega)$, which means that if $g>g_c \equiv \sqrt{\omega \Omega}$, then $\sqrt{\delta_\epsilon}$ is purely imaginary and can no longer be interpreted as an energy gap as in the normal phase. In fact, $\sqrt{\delta_\epsilon}$ can be interpreted as the frequency of a harmonic oscillator that inverts in the superradiant phase~\cite{invertedoscillator2021} [see Eq.~\ref{eq:invosc}]. In the superradiant regime, we then heuristically propose the extension to the above result as
\begin{equation}
\begin{split}
     \hat h_\lambda \simeq &\; \hat H_1 t +\frac{\exp(\sqrt{|\delta_\epsilon|}t)/2-1}{|\delta_\epsilon|}\hat C \\ &-\frac{\exp(\sqrt{|\delta_\epsilon|}t)/2-\sqrt{|\delta_\epsilon|} t}{\sqrt{|\delta_\epsilon|}\delta_\epsilon}\hat D,
     \end{split}
\end{equation}
which means that the quantum Fisher information grows now exponentially with increasing $\sqrt{|\delta_\epsilon|}t$ as we argued before (see Appendix \ref{app:localgenerator} for details). We again contrast with the recently obtained $t^4$ scaling in the normal phase~\cite{felicetti2020criticalqm}. To summarise, at $g=g_c$ one obtains quantum Fisher information whose growth is dictated by photons growing quadratically in time and requires very long  protocol time, similarly to the adiabatic approach to critical quantum metrology~\cite{Rams2018atthelimitsofcriticality,2021gietka_adiabatic}. However, as the coupling parameter becomes significantly larger than $g_c$, the mechanism responsible for the growth of quantum Fisher information becomes the exponential growth of the number of particles caused by squeezing.

\begin{figure*}[htb!]
    \centering
    \includegraphics[width=\textwidth]{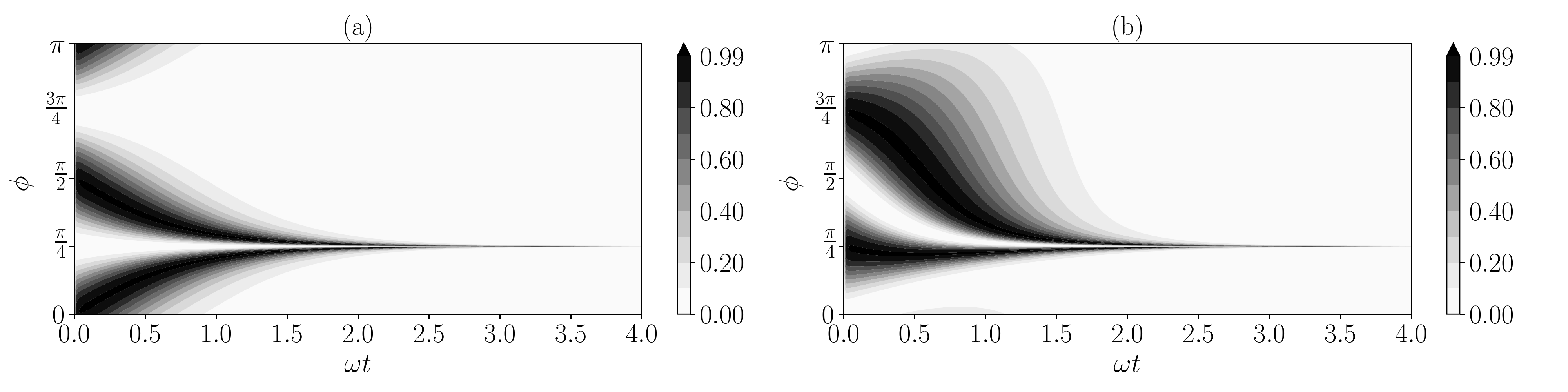}
    \caption{Classical Fisher information for quadrature measurement as a function of quadrature direction $\phi$ and time expressed in units of $\omega$ normalized to the quantum Fisher information. The calculations are performed for $g/g_c \approx \sqrt{2}$ which corresponds to the generation of a perfectly squeezed vacuum in Eq.~\eqref{eq:invertedoscillatorH}. We plot $0<\phi<\pi$ as the information is invariant under rotation by $\pi$. Panel (a) treats $\omega$ as the unknown parameter and panel (b) treats $\Omega$ as the unknown parameter. In both cases the optimal quadrature direction converges to $\pi/4$.}
    \label{fig:fig2}
\end{figure*}

The analytical results showing the growth of the quantum Fisher information as a function of the coupling parameter and time in the superradiant regime are presented in Fig.~\ref{fig:fig1} and Fig.~\ref{fig:fig1b}. For the sake of clarity, we take the logarithm of the normalized quantum Fisher information. Panel (a) of Fig.~\ref{fig:fig1} and Fig.~\ref{fig:fig1b} depict quantum Fisher information treating $\omega$ as the unknown parameter, and (b) depict quantum Fisher information for $\Omega$. One can clearly see that quenching beyond the critical point (the dashed-black line) gives rise to much greater quantum Fisher information than driving in the vicinity of the critical point. This happens because the nonclassical correlations (squeezing) and corresponding excitations (photons) are generated much faster far beyond the critical point. 

Finally, let us discuss the exponential dependence on time of the quantum Fisher information and its relation to the standard quantum limit. In order to derive the latter, one assumes that the initial state containing $N$ particles (atoms or photons) can be prepared instantaneously. In this sense, the standard quantum limit does not include the time and energy needed to prepare the state. Subsequently, the initial state is coherently evolved in time with a time evolution operator that depends on the unknown parameter (unknown parameter imprint) which leads to quadratic dependence on time of the quantum Fisher information (and consequently of the standard quantum limit). In the protocol studied in this work, these two stages---preparation of the state and unknown parameter imprint---happen at the same time. In this sense, it is difficult to make a straightforward and meaningful comparison between the two approaches. One can then ask how to properly quantify quantum enhancement in the critical metrology approach. From the classical point of view, one can clearly identify a quantum enhancement with respect to a classical strategy since lowering the frequency of a harmonic oscillator or even flipping the oscillator upside down does not change the state of the system. A thorough answer to the issue of properly quantifying critical quantum enhancement with respect to existing protocols, however, reaches beyond the scope of this work.


\section{Quadrature measurements}
Although the quantum Fisher information shows exponential growth as a function of time, it does not indicate the measurement that saturates the Cram\'er-Rao bound. As discussed in the introduction, the squeezed vacuum state is the optimal state in single-mode frequency estimation for which the Cram\'er-Rao bound can be saturated using quadrature measurements, which can be realized by a homodyne detection scheme~\cite{2006optimalgaussian,2019homodyneoptimal}. Nevertheless, in general, the resulting state of the considered protocol is not a perfect squeezed vacuum state but a state only resembling the squeezed vacuum. The perfect squeezed vacuum is only generated for $\omega = g^2/2\Omega$, i.e., $g = \sqrt{2 \omega \Omega} = \sqrt{2}g_c$, so when the first term of the Hamiltonian \eqref{eq:invertedoscillatorH} vanishes. In order to determine the optimality of the homodyne detection scheme, we calculate the classical Fisher information
\begin{equation}\label{eq:cfiquadrature}
    \mathcal{F}_\lambda = \sum_\xi \frac{1}{p(\xi|\lambda)}\left(\frac{\partial p(\xi|\lambda)}{\partial \lambda} \right)^2,
\end{equation}
for the distribution obtained by quadrature measurement, where $p(\xi|\lambda)$ is the probability of measuring an outcome labeled by $\xi$ for a given value of an unknown parameter $\lambda$. The homodyne detection boils down to the measurement of a generalized quadrature operator $\hat Q(\phi) = (\hat a e^{i\phi}+ \hat a^\dagger e^{-i\phi})/\sqrt{2}$ which requires additional optimization over the direction  $\phi$; this optimal direction will, in general, depend on the unknown parameter itself (see Appendix \ref{app:variousg}). The dependence of the optimal quadrature direction on time and other parameters of the system is a complication with respect to adiabatic protocols where the optimal direction is always fixed~\cite{felicetti2020criticalqm}. In order to saturate the Cram\'er-Rao bound one could therefore use adaptive real-time feedback techniques~\cite{2015feedbackadaptive}. On the other hand, one could first estimate an approximate value of an unknown parameter with some alternative techniques, and subsequently use an adaptive metrological scheme to maximize the sensitivity~\cite{wiseman_milburn_2009,2011Adaptivemetrology,2017Adaptivemetrologymarkovian,PRXQuantum.2.040301}. In the adaptive approach, the coupling strength $g$ could be increased from relatively low value (even below the critical value) to estimate and update the information about the unknown parameter in every round which should still lead to the possibility of exploiting enhanced rate of creating excitations.

To simplify the cumbersome expression for the classical Fisher information, we first note that the resulting state is Gaussian as the Hamiltonian is quadratic in $\hat a$ and $\hat a^\dagger$. This means that all the information about the state is contained in the mean and variance of quadrature measurements on the wavefunction. As the initial wavefunction has a mean value equal to 0 and is subsequently being squeezed without any displacement, all the information is then stored in the variance~\cite{2018_afr_nek_gaussianstateesitmation,2020Bakmou_gaussian}. As a consequence, measuring the second moment of the quadrature and using it as an estimator allows us to replace the classical Fisher information formula from Eq.~\eqref{eq:cfiquadrature} by the inverse of the error propagation formula (signal-to-noise ratio)~\cite{2012Gaussianreview,2006optimalgaussian}
\begin{equation}
    \mathcal{F}_\lambda = \frac{|\partial_\lambda \hat Q^2(\phi)|^2}{\Delta^2\hat Q^2(\phi)},
\end{equation}
which can be calculated analytically (see Appendix~\ref{app:signaltonoise}). 

\begin{figure}[htb!]
    \centering
    \includegraphics[width=0.5\textwidth]{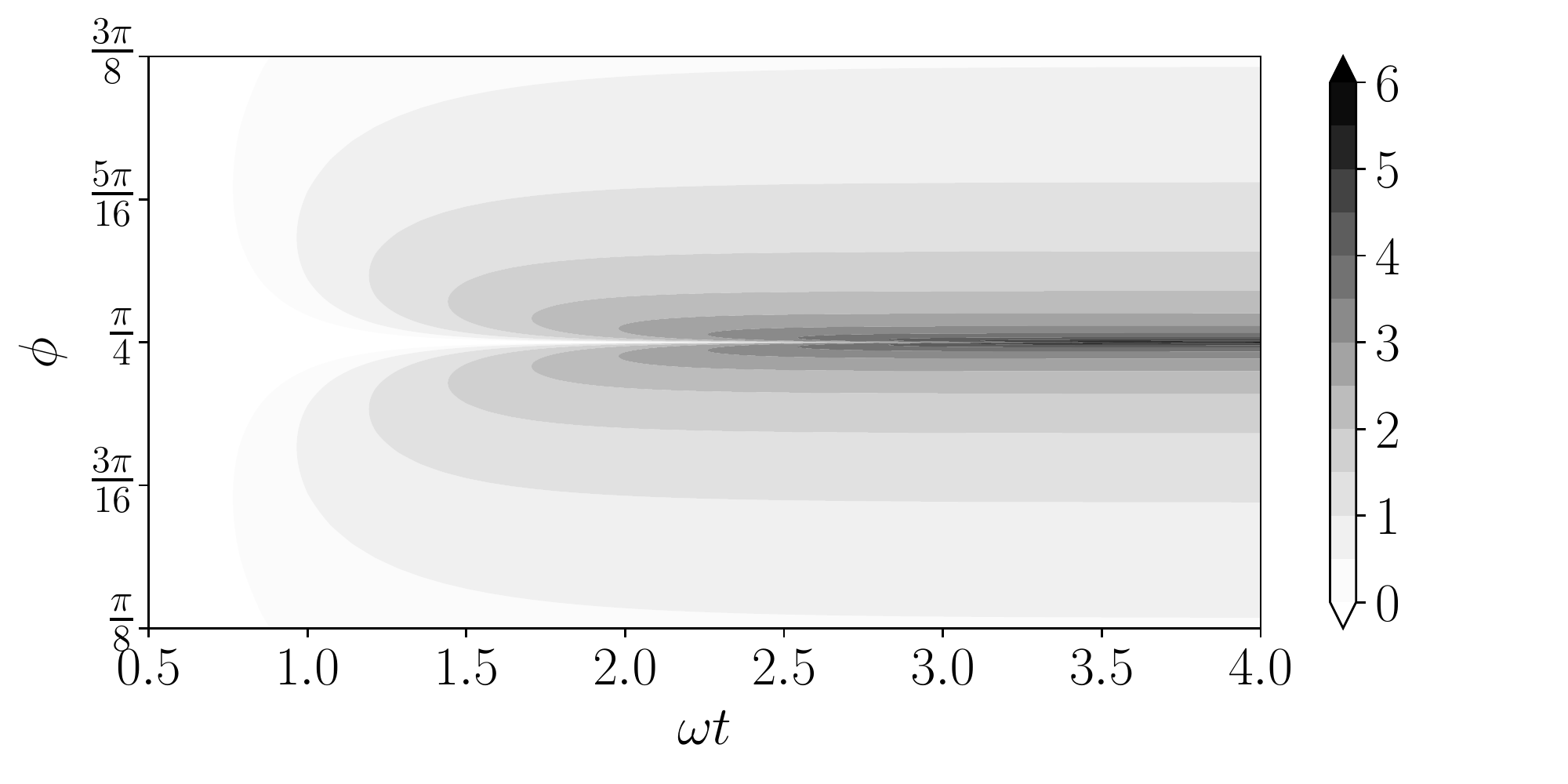}
    \caption{Logarithm of the classical Fisher information from Eq.~\eqref{eq:asymptoticcfi} (multiplied by $\omega^2$). As the classical Fisher information grows, the angle-window for the optimal quadrature direction becomes smaller and converges to a single point (here $\pi/4$). Result for $g = \sqrt{2}g_c$.}
    \label{fig:fig3}
\end{figure}

The analytical results comparing the quantum and classical Fisher information for quadrature measurement as a function of quadrature angle $\phi$ and time for $g = \sqrt{2}g_c$ (calculation for different values of $g$ are given in Appendix~\ref{app:variousg}) are shown in Fig.~\ref{fig:fig2}. As the state is being more squeezed and more photons are generated, the optimal quadrature direction converges exponentially with time to $\pi/4$ which is the direction in which the state is being antisqueezed (for other choices of $g$ the direction of squeezing will be different). This can be explicitly seen if we calculate the classical Fisher information (treating $\omega$ as the unknown parameter)
{\small
\begin{equation}
     \frac{|\partial_\omega \hat Q^2(\phi)|^2}{\Delta^2\hat Q^2(\phi)} =  \frac{2 \cos ^2(2 \phi ) \sinh ^4\left( \omega t \right)}{\omega^2 \left(\cosh \left(2  \omega t\right)-\sin (2 \phi ) \sinh \left(2  \omega t\right)\right)^2},
\end{equation}}
\normalsize
and assuming $\omega t \gg 1$ we obtain to lowest order
\begin{equation}
 \frac{|\partial_\omega \hat Q^2(\phi)|^2}{\Delta^2\hat Q^2(\phi)}  \simeq \frac{\cos ^2(2 \phi )}{2 \omega ^2 (\sin (2 \phi )-1)^2},
\end{equation}
which can be expanded again to lowest order around $\phi = \pi/4$:
\begin{equation}\label{eq:asymptoticcfi}
 \frac{|\partial_\omega \hat Q^2(\phi)|^2}{\Delta^2\hat Q^2(\phi)}  \simeq\frac{1}{2 \omega ^2 \left(\phi -\frac{\pi }{4}\right)^2}.
\end{equation}
This profound result indicates that in order to observe an arbitrary large sensitivity the quadrature direction (phase of the local oscillator) has to be precisely controlled. This can be interpreted as an inherent feature of precision measurements; namely, arbitrarily large sensitives also require an arbitrary precise measuring apparatus. Any uncertainty of such an apparatus---in this case, a potential instability of the measured quadrature angle---will directly influence the precision of estimation (see Fig.~\ref{fig:fig3}).

\section{Dynamical Dicke superradiance in the cavity}
In this section, we briefly discuss the possible implementation of the proposed protocol by exploiting Dicke superradiance in atoms interacting with single-mode cavity field. It is a well-known fact that the self-organization phase transition, which can be observed in systems of thermal~\cite{2003VuleticSELFORG} and ultracold atoms~\cite{baumann2010dicke} trapped and interacting with a cavity mode, can be mapped to the Dicke model~\cite{2010Domokos_Dickemodel}. For simplicity, we consider a one-dimensional zero-temperature Bose-Einstein condensate composed of $N$ atoms each with mass $m$ trapped inside a high-$Q$ single-mode cavity with frequency $\omega_c$ and cavity mode function $\cos kx$ with $x$ being the cavity axis and $k=2\pi \omega_c/c$. The atoms are illuminated from a transverse direction to the cavity axis by a pump laser with frequency $\omega$ far red-detuned from the atomic frequency $\omega_A$. This laser changes the effective photon energy in the cavity (detuning) to $\Delta_c = \omega - \omega_c$. We assume now that the detuning $\Delta_A = \omega - \omega_A$ is much larger than the rate of spontaneous emission such that the atomic excited state can be adiabatically eliminated. The dispersive atom-field interaction strength is $U_0=g_0^2/\Delta_A$ with a single photon Rabi frequency $g_0$, and the effective cavity pump coupling strength describing atom-photon coupling is $\eta = \Omega g_0/\Delta_A$, where $\Omega$ is the Rabi frequency of the coupling to the transverse driving field. Assuming the positional wavefunction is spanned by two Fourier modes $\hat c_0$ and $\hat c_1$ obeying the standard bosonic commutation relations, the Hamiltonian can be written as~\cite{2010Domokos_Dickemodel}
\begin{equation}
\begin{split}
    \hat H_\mathrm{cav} =& -\delta_c\hat a^\dagger \hat a + \omega_R \hat S_z + \frac{i y}{\sqrt{N}} (\hat a^\dagger - \hat a)\hat S_x \\& + u \hat a^\dagger \hat a\left(\frac{1}{2}+\frac{\hat S_z}{N}\right),
    \end{split}
\end{equation}
where $\delta_c = \Delta_c-2u$, $\omega_R = \hbar k^2/2m$, $u = {N U_0}/{4}$, and $y = \sqrt{2 N}\eta$. In the first line of the above Equation we identify the Dicke model, while the second line represents the additional cavity frequency shift inherent to the Bose-Einstein condensate cavity system. The latter, however, does not change the critical coupling~\cite{2010Domokos_Dickemodel} and can be safely neglected in typical experimental situations when $|u|< |\delta_c|$. Applying the Wolff-Schrieffer transformation $\hat U = \exp( y (\hat a^\dagger  -\hat a)\hat S_y/\omega_R\sqrt{N})$ to the above Equation and projecting onto the ground state for $y = 0$ yields (see Appendix~\ref{app:cavity})
\begin{equation}
    \hat H_\mathrm{cav} \simeq \left(-\delta_c + \frac{ y^2}{2  \omega_R}\right) \hat a^\dagger \hat a - \frac{  y^2}{ 4 \omega_R}\left( \hat a^{\dagger2} + \hat a^2 \right),
\end{equation}
which describes a harmonic oscillator with frequency $\nu = 4\delta_c \sqrt{1- (y/y_c)^2}$ and critical coupling $y_c = \sqrt{- \omega_R \delta_c}$. Since the frequency depends on the number of atoms $N$ through $y = \sqrt{2 N}\eta$, the quantum Fisher information picks up an additional power $\sqrt{N}$-enhancement with respect to Eq.~\eqref{eq:QFIdick}, i.e., [assuming $(y/y_c)^2\gg1$]
\begin{equation}
    \mathcal{I}_\lambda \propto \exp\left(8\sqrt{2} \sqrt{\frac{N  |\delta_C |}{\omega_R }}\eta t\right) .
\end{equation} 
This happens because more atoms in the cavity increase the probability of photon scattering into the cavity mode and the number of scattered photons is proportional to $N^2$~\cite{2002HelmutSelforganization}.

From the viewpoint of experimental realizations of the Dicke model in the cavity systems~\cite{Klinder2015dynamics,baumann2010dicke,2014dickemodelraman,2018spinorselforderinglev,2021Kesslertimecrystal}, even though the condition from Eq.~\eqref{eq:initialstatecondition} may not be always satisfied, and the observation of squeezing might not be possible due to the spontaneous symmetry breaking, the exponential growth of the Fisher information should still be an observable effect. This is due to the fact that the exponential growth of the number of photons following the dynamical phase transition (superradiance) occurs for a broad spectrum of parameters even in the mean-field approach (see the experimental results from Ref.~\cite{Klinder2015dynamics}). In a general case, the analytical description of the system might be more obscure than the elegant formalism with an inverted harmonic oscillator as the spin dynamics will not be frozen; nevertheless, the information about the unknown parameter should be still imprinted in the number of photons that grows exponentially in time~\cite{Klinder2015dynamics}. It is worth noting that we have not taken into account the effect of photon losses. Nevertheless, as was shown in Ref.~\cite{invertedoscillator2021}, photon losses will only affect the rate at which the photons are being created and will not destroy metrologically useful correlations stored in a squeezed state. This happens because the squeezed vacuum state is a relatively robust quantum resource. Therefore, even in the presence of decoherence, one can still expect exponential growth of the quantum and classical Fisher information in time.


\section{Conclusions \& Outlook}
In this work, in order to eliminate the critical slowing down that hinders protocols of critical quantum metrology, we have proposed to quench the system far beyond the critical point and thus increase the rate at which correlations are created. The presented idea is significantly faster (i.e., exponential versus power-law scaling in time) than existing adiabatic and dynamical protocols, which can be of fundamental importance in realistic scenarios where deleterious effects of decoherence and experimental noise are inevitable. The price being paid for the enhancement is an additional dependence of the optimal quadrature angle on the parameters of the system. This however could be overcome by using an adaptive strategies.

The protocol can be understood as a non-linear type of interferometry where the information about the unknown parameter is not only stored in the phase of the wave function but also in the number of excitations (photons) and the nonclassicality of them itself. The results are fully valid close to the thermodynamic limit but can also be applied to finite size systems under additional constraints on time~\cite{invertedoscillator2021}. They can also be extended to a more general case in which the inverted harmonic oscillator description may not be fully valid. In this case the spin degree of freedom will also experience squeezing and lead to light-matter entanglement.

As quenching the system into the superradiant phase generates a Gaussian state, we have argued and shown analytically that the quadrature measurement can saturate the quantum Cram\'er-Rao bound for an optimal angle outside of the normal phase. The optimal quadrature angle becomes effectively a point in the limit of $\omega t \gg 1$ which can be understood as a condition that a measuring apparatus has to satisfy in order to perform a very precise measurement. 

Our protocol can be readily tested in quantum simulators which realize the (optical) quantum Rabi model \cite{rauschnbeutel2018observationofultrastrong,solano2019strongcouplingreview,kockum2019ultrastrong}, and quantum simulators which can realize the (optical) Dicke model \cite{baumann2010dicke,2014dickemodelraman}, i.e., a macroscopic number of spins coupled to a single-mode field. As discussed in this work, a promising candidate is a gas of atoms in an optical cavity. Moreover, we have shown that these systems can exhibit sensitivity that additionally scales exponentially with the square root of  the number of atoms $N$ confirming that exponentially scaling  Fisher information can be a physical and observable effect~\cite{2008Exponetially_enhanced_quantum_metrology_PRL} and does not necessarily require a $N$-body term. 

The presented idea paves a way towards metrology protocols which can take benefit from storing the information about the unknown parameter in the number of excitations. An intriguing possibility is testing the proposed protocol in quantum simulators which realize the Dicke (Rabi) model without photonic degrees of freedom~\cite{engels2014dickesocbec,2018reydickeions,2019rydbergiondicke,konstantinov2019PhysRevLett.122.176802}. Our analysis shows that measuring simply excitations of the system---such as center-of-mass excitations in a spin-orbit coupled Bose-Einstein condensate~\cite{engels2014dickesocbec}---could lead to enhanced metrology. 


\begin{acknowledgments}
The authors are pleased to acknowledge Farokh Mivehvar, Friederieke Metz, Tim Keller, and Jing Li for inspiring discussions and thank the anonymous Referees for their suggestions and comments. K.G. would like to thank Szemek Keppe for his support. Simulations were performed using the open-source QuantumOptics.jl framework in Julia~\cite{kramer2018quantumoptics}. This work was supported by the Okinawa Institute of Science and Technology Graduate University. K.G. acknowledges support from the Japanese Society for the Promotion of Science (P19792).
\end{acknowledgments}
\onecolumngrid
\appendix
\section{Quantum Fisher information for the eigen-states of the Dicke model in the normal phase}\label{app:qfinormal}
\noindent The quantum Fisher information can be calculated by using the following formula
\begin{equation}
    \mathcal{I}_\lambda = 4(\langle \partial_\lambda \psi|\partial_\lambda \psi \rangle - \langle \partial_\lambda \psi|\psi \rangle^2),
\end{equation}
where $\lambda$ is an unknown parameter. The $n$th eigenstate $|\psi_n \rangle$ for the Dicke model in the thermodynamic limit below the critical point is
\begin{equation}
|\psi_n \rangle = \hat S(\xi)| n \rangle \otimes |\!\downarrow \,\rangle,
\end{equation}
where $\hat S(\xi) \equiv \exp\{(\xi/2)(\hat a^\dagger)^2-(\xi^*/2)\hat a^2\}$ is the squeezing operator with $\xi = -\frac{1}{4} \ln\{1-(g/g_c)^2\}$ and $g_c = \sqrt{\omega\Omega}$. We begin by calculating the derivative of the instantaneous eigenstate $|\psi_n \rangle$ with respect to an unknown parameter which we assume to be $\omega$. This yields
\begin{equation}
| \partial_\omega \psi_n\rangle =\frac{g^2}{8 \omega (g^2-\omega\Omega)}\left[\left(\hat{a}^\dagger\right)^2 - \hat{a}^2\right]  \hat S(\xi) | n \rangle \otimes |\!\downarrow \,\rangle,
\end{equation}
and allows to calculate the overlap $\langle \partial_\omega \psi_n|\psi_n \rangle$
\begin{equation}
\begin{split}
    \langle \partial_\lambda \psi_n|\psi_n \rangle &= \frac{g^2}{8 \omega (g^2-\omega\Omega)} \langle n | \hat S^\dagger(\xi) \left[\left(\hat{a}^\dagger\right)^2 - \hat{a}^2\right] \hat  S(\xi)|n\rangle  \\ & = \frac{g^2}{8 \omega (g^2-\omega\Omega)} \langle n | \left[\left(\hat{a}^\dagger \cosh(|\xi|) + \hat{a} \sinh(|\xi|)\right)^2 -\left(\hat{a}\cosh(|\xi|) + \hat a^\dagger \sinh(|\xi|)\right)^2 \right] |n\rangle \\ & = 
    \frac{g^2}{8 \omega (g^2-\omega\Omega)}  \cosh(|\xi|)\sinh(|\xi|)  \langle n | \left(\hat{a}^\dagger \hat a+ \hat a \hat a^\dagger - \hat a^\dagger \hat a - \hat a \hat a^\dagger\right) |n\rangle = 0,
    \end{split}
\end{equation}
where we have used
\begin{align}
\begin{split}
\hat S^\dagger(\xi)\hat a \hat S(\xi) = \hat{a}\cosh(|\xi|) - e^{i\arg(\xi)} \hat a^\dagger \sinh(|\xi|) = \hat{a}\cosh(|\xi|) + \hat a^\dagger \sinh(|\xi|).
\end{split}
\end{align}
An analogous (but much more tedious) calculation for the $\langle \partial_\omega \psi_n| \partial_\omega\psi_n \rangle$ yields 
\begin{equation}
    \langle \partial_\omega \psi_n| \partial_\omega\psi_n \rangle =\frac{g ^4 (1 + n + n^2)}{32 \omega^2 \left(g ^2-\omega \Omega \right)^2}.
\end{equation}
Substituting the $\langle \partial_\omega \psi_n|\psi_n \rangle$ and $\langle \partial_\omega \psi_n| \partial_\omega\psi_n \rangle$ into the quantum Fisher information formula gives
\begin{equation}
    \mathcal{I}_\omega = \frac{g^4 (1 + n + n^2)}{8 \omega^2 \left(g ^2-\omega \Omega \right)^2} =  \frac{ (1 + n + n^2)}{8 \omega^2  \left(1 - \frac{g ^2}{g_c^2} \right)^2} \frac{g ^4}{g_c^4}.
\end{equation}
Treating $\Omega$ as an unknown parameter one obtains
\begin{equation}
    \mathcal{I}_\Omega  =  \frac{ (1 + n + n^2)}{8 \Omega^2  \left(1 - \frac{g ^2}{g_c^2} \right)^2} \frac{g ^4}{g_c^4}.
\end{equation}
This extends the result of $n=0$~\cite{felicetti2020criticalqm} to an arbitrary $n$. 

{\section{Schrieffer-Wolff transformation of the Dicke model}\label{app:guidline}
In the case of the Dicke model
\begin{equation}
    \hat H_{\mathrm{DM}} = \omega \hat a^\dagger \hat a + {\Omega} \hat S_z +\frac{g}{\sqrt{N}}\left( \hat a^\dagger + \hat a \right) \hat S_x,
\end{equation}
the Schrieffer-Wolff transformation is a simultaneous rotation of the (collective) spin around the $y$-axis and displacement of the harmonic oscillator 
\begin{equation}
 \hat U_{\mathrm{SW}} = \exp\big\{i \frac{g}{g_c} \frac{\sqrt{\omega}}{\sqrt{\Omega N}} \left(\hat a^\dagger + \hat a\right){\hat S_y}\big\}.
\end{equation}
As a result of the transformation, one obtains
\begin{align}
\begin{split}
  \hat H_{\mathrm{DM}} =&\,\, \omega \left(\hat a^\dagger +i \frac{g}{g_c}\frac{\sqrt{\omega}}{\sqrt{\Omega N }} \hat S_y\right) \left(\hat a -i\frac{g}{g_c}\frac{\sqrt{\omega}}{\sqrt{\Omega N}} \hat S_y\right) \\&+ {\Omega}\left(\cos\left(\frac{g}{g_c}\frac{\sqrt{\omega}}{\sqrt{\Omega N}}(\hat a^\dagger + \hat a)\right)\hat S_z - \sin\left(\frac{g}{g_c}\frac{\sqrt{\omega}}{\sqrt{\Omega N}}(\hat a^\dagger + \hat a)\right)\hat S_x\right)\\
  &+ \frac{g}{\sqrt{N}}\left(\hat a^\dagger+ \hat a \right)\left(\cos\left(\frac{g}{g_c}\frac{\sqrt{\omega}}{\sqrt{\Omega N}}(\hat a^\dagger + \hat a)\right)\hat S_x + \sin\left(\frac{g}{g_c}\frac{\sqrt{\omega}}{\sqrt{\Omega N}}(\hat a^\dagger + \hat a)\right)\hat S_z\right).
  \end{split}
\end{align}
Expanding the above Hamiltonian in the Taylor series assuming $ \omega/\Omega \ll 1$ gives Eq.~\eqref{eq:effqr} from the main text
\begin{equation}
  \hat H_{\mathrm{DM}} =\left(\omega + \frac{g^2}{\Omega N}\hat S_z \right) \hat a^\dagger \hat a + {\Omega} \hat S_z +\frac{g^2}{2\Omega N}\left( \hat a^{\dagger2} + \hat a^2 \right) \hat S_z.
\end{equation}
If the initial state is a collective spin down state, i.e. $\hat S_z |\!\downarrow\,\rangle = -N/2|\!\downarrow\,\rangle$, then the Hamiltonian can be simplified to
\begin{equation}
  \hat H_{\mathrm{DM}} =\left(\omega - \frac{g^2}{\Omega 2} \right) \hat a^\dagger \hat a +\frac{g^2}{4\Omega }\left( \hat a^{\dagger2} + \hat a^2 \right),
\end{equation}
which can be rewritten in the form of a harmonic oscillator with a tunable frequency
\begin{equation}
   \hat H = \frac{\omega}{2} \hat P^2 + \frac{\omega}{2}\left(1-\frac{g^2}{g_c^2}\right)\hat X^2.
\end{equation}
The above Hamiltonian can be used to calculate the ground state but only in the normal phase. From the viewpoint of metrology, however, it is more interesting to explore the dynamics of this Hamiltonian. Following Ref. \cite{invertedoscillator2021}, it can be shown that this Hamiltonian gives rise to proper (squeezing) dynamics until certain number of excitations is reached. To be more precise this number can be calculated to be
\begin{equation}
    \left\langle \hat a^\dagger \hat a \right\rangle_t < \frac{1}{32} {\frac{N \Omega }{ \omega }}\left({\frac{g^2}{g_c^2}-\frac{g_c^2}{g^2}}\right),
\end{equation}
which can be further elevated by increasing $N$.
}

\section{Local generator for the Dicke model under the Schrieffer-Wolff transformation}\label{app:localgenerator}
Following the derivation from Ref.~\cite{chu2021dynamiccriticalmetrology}, it can be shown that the transformed local generator $\hat h_\lambda$ for a Hamiltonian $\hat H = \hat H_0 +\lambda \hat H_\lambda$ can be expressed as
\begin{equation}
     \begin{split}
       \hat h_\lambda =  i e^{i \hat H t} \left(\partial_\lambda e^{-i \hat  H t}\right) = \int_0^t   e^{i \hat H s} \hat H_\lambda  e^{-i \hat  H s} \mathrm{d}s  =  \int_0^t \sum_{n=0}^\infty \frac{(is)^n}{n!}\left[\hat H,\hat H_\lambda \right]_n \mathrm{d}s
       =  - i \sum_{n=0}^\infty \frac{(it)^{n+1}}{(n+1)!}\left[\hat H,\hat H_\lambda \right]_n,
     \end{split}
 \end{equation}
 where $[\hat H,\hat H_\lambda ]_{n+1} =[\hat H, [\hat H,\hat H_\lambda ]]_{n}$ and $[\hat H,\hat H_\lambda]_{0} = \hat H_\lambda$. Now it is possible to show that $[\hat H,\hat H_\lambda] = i \hat C$, $[\hat H, i \hat C] = -\hat D$, and $[\hat H,- \hat D] = i \delta_\epsilon \hat C$, which results in the following commutation relations
 \begin{equation}
         [\hat H,\hat H_\lambda ]_{2n+1} = i \delta_\epsilon^n \hat C \quad \mathrm{ and } \quad
         [\hat H,\hat H_\lambda ]_{2n+2} = - \delta_\epsilon^n \hat D. 
 \end{equation}
Therefore, one can express the local generator $\hat h_\lambda$ as
\begin{equation}
\hat h_\lambda = \hat H_\lambda t - i \sum_{n=0}^\infty \frac{(it)^{2n+2}}{(2n+2)!}i\delta_\epsilon^n \hat C + i \sum_{n=0}^\infty \frac{(it)^{2n+3}}{(2n+3)!}\delta_\epsilon^n \hat D =\hat H_\lambda t +\frac{\cos(\sqrt{\delta_\epsilon}t)-1}{\delta_\epsilon}\hat C -\frac{\sin(\sqrt{\delta_\epsilon}t)-\sqrt{\delta_\epsilon} t}{\sqrt{\delta_\epsilon}\delta_\epsilon}\hat D.
\end{equation}
In the low-energy sector we can rewrite the Dicke model as
\begin{equation}
    \hat H = \frac{\omega}{2}\left(\hat P^2 +\hat X^2\right)- \frac{g^2}{2\Omega}\hat X^2,
\end{equation}
with $\hat X = (\hat a + \hat a^\dagger)/\sqrt{2}$ and $\hat P = (\hat a - \hat a^\dagger)/\sqrt{2}i$. Assuming that $\omega$ is unknown, by choosing $\hat H_\omega = \frac{1}{2}(\hat X^2+\hat P^2)$ and $\hat H_0 = -\frac{g^2}{2\Omega} \hat X^2$, one can easily show that
\begin{equation}
\hat C = -\frac{g^2}{2\Omega}\left(\hat X \hat P + \hat P \hat X \right), \quad
\hat D = -\frac{g^2 \omega}{\Omega}(\hat X^2-\hat P^2) +\frac{g^4}{\Omega^2}\hat X^2, \quad \mathrm{and}\quad
\delta_\epsilon =  4\omega^2\left(1-\frac{g^2}{\omega \Omega}\right),
\end{equation}
leading to
\begin{equation}
\begin{split}
\hat h_\omega  = &\frac{t}{2}\left(\hat X^2+\hat P^2\right) -\frac{g^2}{2g_c^2 \omega}\frac{\cos\left(\sqrt{\delta}\omega t\right)-1}{ \delta}\left(\hat X \hat P + \hat P \hat X\right) \\
&+\frac{g^2 }{g^2_c \omega}\frac{\sin\left(\sqrt{\delta} \omega t\right)-\sqrt{\delta} \omega t}{\sqrt{\delta}\delta}\left(\hat X^2-\hat P^2\right)  - \frac{g^4}{g_c^4\omega}\frac{\sin\left(\sqrt{\delta} \omega t\right)-\sqrt{\delta} \omega t}{\sqrt{\delta}\delta}\hat X^2,
\end{split}
\end{equation}
where $\delta = \delta_\epsilon/\omega^2 = 4(1-\frac{g^2}{\omega \Omega})$ and $g_c =\sqrt{\omega \Omega}$. Analogous calculations can be performed treating $\Omega$ as an unknown parameter yielding
\begin{equation}
\begin{split}
\hat h_\Omega  = & \frac{g^2 \omega t}{2 g_c^2 \Omega} \hat X^2 +\frac{g^2}{2 g_c^2 \Omega}\frac{\cos\left(\sqrt{\delta}\omega t\right)-1}{ \delta}\left(\hat X \hat P + \hat P \hat X\right) \\
&+\frac{g^2 }{ g_c^2 \Omega}\frac{\sin\left(\sqrt{\delta} \omega t\right)-\sqrt{\delta} \omega t}{\sqrt{\delta}\delta}\left(\hat X^2-\hat P^2\right)  - \frac{g^4}{g_c^4\Omega}\frac{\sin\left(\sqrt{\delta} \omega t\right)-\sqrt{\delta} \omega t}{\sqrt{\delta}\delta}\hat X^2.
\end{split}
\end{equation}
Similar local generator can be found for $g$ as has been shown in Ref.~\cite{chu2021dynamiccriticalmetrology}
\begin{equation}
\begin{split}
    \hat h_g = &-\frac{g\sqrt{\omega} t}{g_c\sqrt{\Omega}}\hat X^2 -\frac{ g}{ g_c^2} \frac{\cos\left(\sqrt{\delta}\omega t\right)-1}{ \delta}\left(\hat X \hat P + \hat P \hat X\right) \\
    & -\frac{ 2g}{g_c^2}\frac{\sin\left(\sqrt{\delta} \omega t\right)-\sqrt{\delta} \omega t}{\sqrt{\delta}\delta}\left(\hat X^2-\hat P^2\right)  + \frac{2g^3}{g_c^4}\frac{\sin\left(\sqrt{\delta} \omega t\right)-\sqrt{\delta} \omega t}{\sqrt{\delta}\delta}\hat X^2.
\end{split}
\end{equation}

\section{Error propagation formula for the second moment of Quadrature estimation}\label{app:signaltonoise}
In order to calculate the error propagation formula, we first move to the Heisenberg picture and evaluate how $\hat a$ and $\hat a^\dagger$ evolve in time. Using matrix notation it can be easily shown that a Bogoliubov transformation is obtained:
\begin{equation}
      \begin{pmatrix} \hat a(t) \\ \hat a^\dagger(t) \end{pmatrix} = \exp\left(i t \hat A\right) \begin{pmatrix} \hat a \\ \hat a^\dagger  \end{pmatrix},
\end{equation}
where 
\begin{equation}
    \hat A = \begin{pmatrix}
 -\frac{g^2}{2 \Omega } & \frac{g^2}{2 \Omega }-\omega \\
 \omega-\frac{g^2}{2 \Omega } & \frac{g^2}{2 \Omega }  \end{pmatrix}.
\end{equation}
Upon diagonalization of the above matrix, one can find
\begin{equation}
\begin{split}
    \hat a(t) = \hat a \cos \left(\sqrt{\delta} \omega t/2 \right) - \hat a \frac{2 i   \sin \left(\sqrt{\delta} \omega t/2\right)}{\sqrt{\delta}}+ (\hat a - \hat a^\dagger)\frac{i g^2 \sin \left(\sqrt{\delta} \omega t/2\right)}{ g_c^2  \sqrt{\delta}},
        \end{split}
\end{equation}
and similarly for $\hat a^\dagger(t)$. The second moment of the quadrature operator for initial vacuum state can now be calculated to give
\begin{equation}
    \begin{split}
        \langle 0| \hat Q^2(t) | 0 \rangle =& \frac{1}{2} -\frac{g^2 \sin (2 \phi )  \sin \left( \sqrt{\delta} \omega t\right)}{2 g_c^2  \sqrt{ \delta}}+\frac{g^2 \cos (2 \phi ) \cos \left( \sqrt{\delta} \omega t\right)}{ g_c^2\delta} -\frac{g^2 \cos (2 \phi )}{2 g_c^2\delta}\\&+\frac{g^4 \sin ^2\left(\sqrt{\delta} \omega t/2\right)}{  g_c^4\delta}+\frac{g^4 \cos (2 \phi ) \sin ^2\left(\sqrt{\delta} \omega t/2\right)}{2  g_c^4\delta},
    \end{split}
\end{equation}
and all the higher moments can be calculated from this expression as the distribution is Gaussian. In order to get a further insight, we can now focus on the special case of $g = \sqrt{2}g_c$, which boils down to pure squeezing, and set $\omega$ as the unknown parameter (without the loss of generality). Evaluating the signal-to-noise ratio and substituting $g = \sqrt{2}g_c$ into the expression yields
\begin{equation}
  \frac{|\partial_\omega \hat Q^2(\phi)|^2}{\Delta^2\hat Q^2(\phi)} =  \frac{2 \cos ^2(2 \phi ) \sinh ^4\left( \omega t \right)}{\omega^2 \left(\cosh \left(2  \omega t\right)-\sin (2 \phi ) \sinh \left(2  \omega t\right)\right)^2}.
\end{equation}
Assuming now $\omega t \gg 1$ such that $\cosh(\omega t) \simeq \sinh(\omega t) \simeq \exp(\omega t)/2$, we finally obtain
\begin{equation}
    \frac{|\partial_\omega \hat Q^2(\phi)|^2}{\Delta^2\hat Q^2(\phi)}  \simeq \frac{\cos ^2(2 \phi )}{2 \omega ^2 (\sin (2 \phi )-1)^2},
\end{equation}
which diverges as $\phi$ approaches $\pi/4$ which can be explicitly seen by expanding in Taylor series around $\phi = \pi/4$
\begin{equation}
 \frac{|\partial_\omega \hat Q^2(\phi)|^2}{\Delta^2\hat Q^2(\phi)}  \simeq \frac{1}{2 \omega ^2 \left(\phi -\frac{\pi }{4}\right)^2}.
\end{equation}

\section{Classical Fisher information for an arbitrary $g/g_c$ in the superradiant phase}\label{app:variousg}
\begin{figure*}[htb!]
    \centering
    \includegraphics[width=1\textwidth]{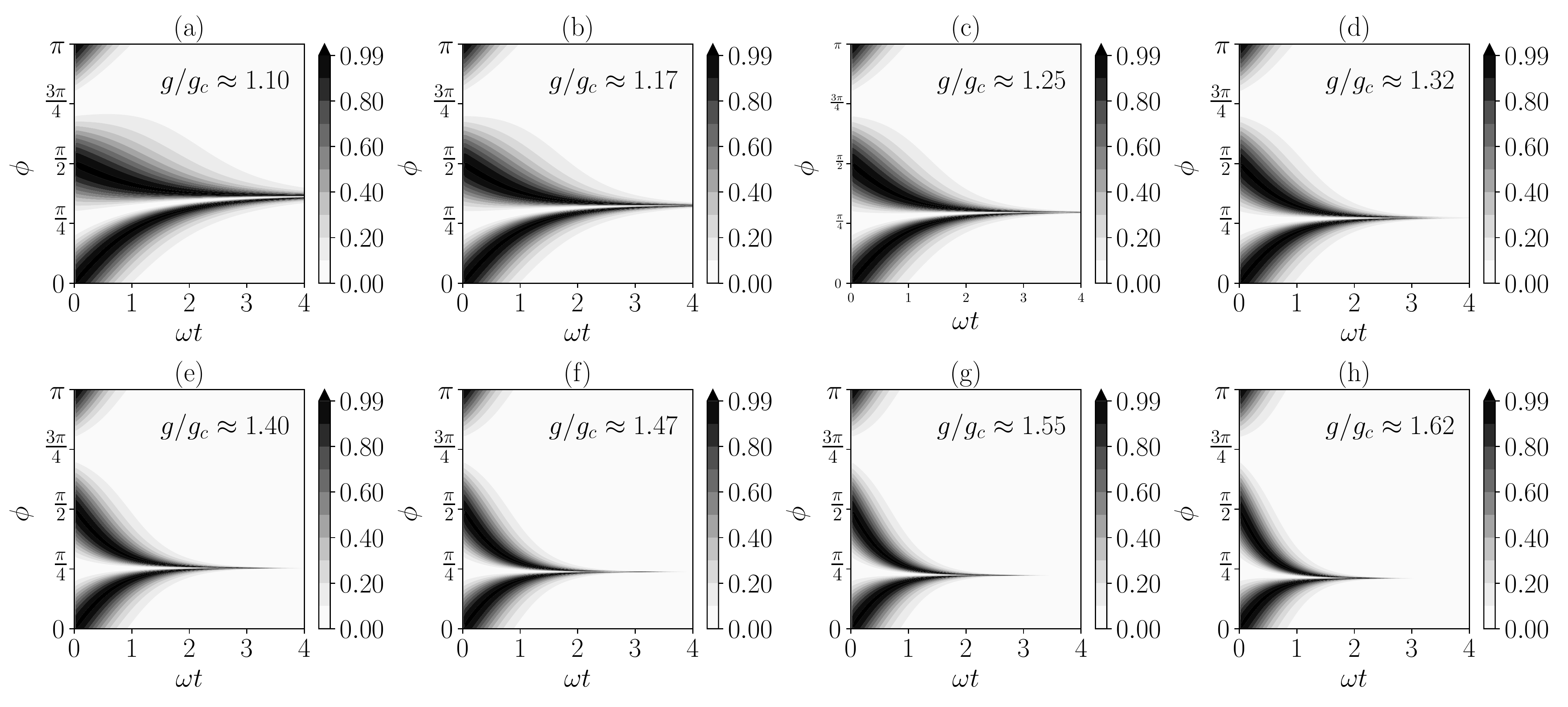}
    \caption{Classical Fisher information normalized to the quantum Fisher information as a function of $\phi$ and time for various values of $g/g_c$. In all the cases the homodyne detection in the optimal direction saturates the Cram\'er-Rao bound.}
    \label{fig:fig4}
\end{figure*}

In the main text, we have presented calculations for a special case of $g = \sqrt{2}g_c$ which corresponds to pure squeezing. The natural question that arises is whether the Cram\'er-Rao bound can be saturated for other choices of $g/g_c$. To this end, let us calculate the optimal angle of the quadrature direction. Assuming $g/g_c = \sqrt{r}$, the optimal direction is given by
\begin{equation}
    \phi = \cos^{-1}\left(\sqrt{\frac{f(r)}{2}} \right),
\end{equation}
with
\begin{equation}
\begin{split}
    f(r) =& \frac{2 r^2 \sinh ^2\left(\sqrt{r-1}  \omega t \right)+8 r-8}{r^2 \cosh \left(2 \sqrt{r-1}  \omega t\right)-r^2+8r-8} \\&+
    \frac{\sqrt{2} \text{csch}^2\left(\sqrt{r-1}  \omega t \right) \sqrt{(r-2)^2 \sinh ^6\left(\sqrt{r-1}  \omega t \right) \left(r^2 \cosh \left(2 \sqrt{r-1}  \omega t \right)-r^2+8r-8\right)}}{r^2 \cosh \left(2 \sqrt{r-1}  \omega t \right)-r^2+8r-8}.
    \end{split}
\end{equation}
In the limit of $ \sqrt{r-1}\omega t\rightarrow\infty$ the optimal angle becomes
\begin{equation}\label{eq:optimalangle}
    \lim_{t \to \infty} \phi = \cos ^{-1}\left({\sqrt{1-\frac{g_c^2}{g^2}}}\right),
\end{equation}
which is the direction perpendicular to the squeezing direction as discussed in the main text. In order to show the saturation of the Cram\'er-Rao bound for other values of $g/g_c$ we performed simulations whose results are presented in Fig.~\ref{fig:fig4}. The figure depicts classical Fisher information normalized to quantum Fisher information as a function of $\phi$ and time. As can be seen, in every case the saturation of the Cram\'er-Rao bound is possible. We interpret this result as a potential confirmation of the saturation of the Cram\'er-Rao bound by the homodyne detection scheme. For $g/g_c>1.25$, the simulations cannot resolve the peak of the classical Fisher information as the optimal quadrature angle converges to a single point given by Eq.~\eqref{eq:optimalangle}.

\section{Schrieffer-Wolff transformation for cavity superradiance}\label{app:cavity}
Following Ref.~\cite{2010Domokos_Dickemodel} a quantum gas of atoms interacting dispersively with a cavity mode can be mapped to the following Hamiltonian
\begin{equation}
    \hat H_\mathrm{cav} = -\delta_C\hat a^\dagger \hat a + \omega_R \hat S_z + \frac{i y }{\sqrt{N}}(\hat a^\dagger - \hat a)\hat S_x + u \hat a^\dagger \hat a\left(\frac{1}{2}+\frac{\hat S_z}{N}\right).
\end{equation}
Using the Schreiffer-Wolff transformation of the form
\begin{equation}
    \hat U = \exp \left( \frac{y}{\omega_R\sqrt{N}}\left(\hat a^\dagger -\hat a\right)\hat S_y \right),
\end{equation}
and assuming $|\delta_C|/\omega_R N \ll 1$, $|u|<|\delta_C|$, neglecting terms proportional to $\sqrt{\delta_c/\omega_R N}$ gives
\begin{equation}
    \hat H_\mathrm{cav} \approx  -\delta_C\hat a^\dagger \hat a + \omega_R \hat S_z + \frac{y^2}{2 \omega_R N}(\hat a^\dagger - \hat a)^2 \hat S_z + u \hat a^\dagger \hat a\left(\frac{1}{2}+\frac{\hat S_z}{N}\right).
\end{equation}\balance
The above Hamiltonian commutes with the $\hat S_z$ operator which means that $\langle \hat S_z\rangle$ is a constant of motion. Therefore, if the initial state is a collective spin-down state, the operator can be replaced by the eigenvalue $-N/2$ to give
\begin{equation}
    \hat H_\mathrm{cav} \approx  -\delta_C\hat a^\dagger \hat a  - \frac{ y^2}{4 \omega_R}(\hat a^\dagger - \hat a)^2.
\end{equation}
The considered transformation can be understood as an elimination (time-dependent) of the spin degree of freedom.
\newline

\end{document}